\documentstyle[prl,twocolumn,aps,epsf]{revtex}

\newcommand{\Va}{NaV$_2$O$_5\;$}

\begin{document}
\tolerance 50000
\draft
\twocolumn[\hsize\textwidth\columnwidth\hsize\csname@twocolumnfalse\endcsname

\title{Exact diagonalisation study of charge order in the quarter-filled\\
two-leg ladder system \Va}

\author{A. Langari$^{1}$, 
M. A. Mart\'{\i}n-Delgado$^{2}$
and P. Thalmeier$^3$ 
}

\address{$^1$Max-Planck-Institute for the Physics of Complex Systems, 
D-01187 Dresden, Germany\\
$^2$Departamento de F\'{\i}sica Te\'orica I, Universidad Complutense,
28040-Madrid, Spain\\
$^3$Max-Planck-Institute for Chemical Physics of Solids\\ D-01187 Dresden, 
Germany}

\date{Feb. 1, 2001}
\maketitle

\begin{abstract}
The charge ordering transition in the layer compound \Va is studied by means
of  exact diagonalization methods for finite systems. 
The 2-leg ladders of the V-Trellis lattice are associated with one spin variable 
of the vanadium 3d-electron in the rung and a pseudospin 
variable that  describes its positional degree of freedom. 
The charge ordering (CO) due to intersite Coulomb interactions is 
described by an effective Ising-like Hamiltonian for the pseudo-spins that are 
coupled to the spin fluctuations along the ladder. We employ a 
Lanczos algortihm on 2D lattice to compute charge (pseudo-spin) and spin-correlation 
functions and the energies of the low lying excited states. 
A CO-phase diagram is constructed and the effect of intra-ladder 
exchange on the CO transition is studied. It is shown that a 
phase with no-longe range order (no-LRO)
exists between the in-line and zig-zag ordered structures. 
We provide a finite-size scaling analysis for the spin excitation gap
and also discuss the type of excitations.  
In addition we studied the effect of bond-alternation of spin exchange and 
derived a scaling form for the spin gap in terms of the dimerization 
parameter.
\begin{center}
\parbox{14cm}{}

\end{center}
\end{abstract}

\vspace{-0.8 true cm}

\pacs{\hspace{2.5cm}
} 
\vskip2pc]

\section{Introduction}

The interplay of charge ordering (CO) and spin structure and 
dynamics has recently 
been in the focus of interest especially in the layered 3d-perovskites like 
cubic and bilayer manganites. A CO transition is caused
by inter-site Coulomb interactions dominating the kinetic hopping energy of 
the electrons of transition elements below a critical temperature T$_c$. It
localizes the electrons and thus defines an associated low temperature spin
lattice which then may exhibit magnetic ordering and possibly low 
dimensional spin excitations.
Such an interplay of charge order and spin excitations, supplemented by
a lattice distortion is present 
in the Trellis lattice compound \Va. Its basic building 
blocks 
consist of two leg ladders whose rungs are occupied by one d-electron on 
the average (Fig.\ref{fig1}). It is an insulating compound with an effective 
charge transfer gap given by the hopping t$_a$ across the rung. The electronic 
structure of this compound was investigated in 
Refs.(\onlinecite{Fuji}-\onlinecite{Horsch}) 
in detail and hopping parameters for a tight binding model were derived.
Susceptibility measurements \cite{Isobe} NMR- experiments \cite{Ohama} and 
x-ray analysis \cite{Luedecke} have shown that a spin gap evolves 
below T$_c= 33 ^{\circ}K$ at the same 
time when \Va undergoes a CO- transition and lattice distortion. Sofar 
there is no complete understanding of both the low temperature structure 
and the origin of the spin gap. 

A model for the 
coupling of CO to spin degrees of freedom in \Va was 
introduced in Ref.(\onlinecite{Thalmeier1}) based on a pseudospin 
representation for 
the rungs such that the CO transition can be described within a 
generalized 2D Ising model. Many proposals for the proper low temperature
structure have been made. According to the model by 
L\"udecke et al \cite{Luedecke} only every second ladder has zig-zag CO 
whereas the other ladder stays disordered. This CO is also 
supported by neutron scattering results \cite{Regnault} discussed in 
Ref.(\onlinecite{Thalmeier2}). 
On the other hand, this charge distribution disagrees with an x-ray 
anomalous scattering measurement which indicates charge modulation along 
the b-axis \cite{nakao}.
Furthermore the structure in Ref.(\onlinecite{Luedecke}) leads to three 
inequivalent V- sites in valence states V$^{4+}$, V$^{4.5+}$ and V$^{5+}$
which is incompatible with 
NMR results \cite{Ohama} which show that only two inequivalent V-sites
exist. 
A solution of this puzzle has been proposed in Ref.(\onlinecite{Bernert}) 
where it was shown that this structure actually leads to asymmetric and 
incomplete charge order where two of the inequivalent sites have almost 
identical valences. However incomplete charge order complicates 
the naive application of conventional exchange models for the spin 
dynamics. Therefore it is first necessary to study the influence of 
charge ordering on the superexchange mechanism.
It was shown \cite{Yushankhai} how charge correlations in the rungs 
and CO progressively reduce the superexchange of spins which occupy rung 
molecular orbitals. 
For this calculation to be carried through, some mean field like 
approximations in the charge (pseudo- spin) variables have to be made to start 
from a restricted orbital space for the calculation of exchange 
integrals.
Finally we mention that some recent results\cite{deBoer} emphasize the 
fully zig-zag CO in the ladders below T$_c$ for \Va, but the stacking 
perpendicular to the ladders for this model is still controversial.

As long as the critical temperature (T$_c$) is finite one may guess that 
only classical fluctuations are responsible for the phase transition phenomena.
However an estimate of the critical point in the whole phase diagram will 
show whether the quantum fluctuations should be considered. Since the
scale of hopping energies and Coulomb interactions are in the order of 
1 eV then the critical temperature in this unit is T$_c$=0.0028 eV.
This shows that our model is very close to quantum critical point and
quantum fluctuations are very important in the phase transition. Moreover the 
results at T=0 can explain the phase transition phenomena to a very good
extent.    
It is therefore desirable to do a fully quantum mechanical 
calculation which treats the charge (pseudo- spin $\vec{\tau}$) 
and spin $\vec{\sigma}$ variables of the V-V rungs on the same 
level. Such an approach is presented in this work and carried 
through by using exact diagonalisation methods with the Lanczos 
algorithm. In Sec. II we introduce the pseudospin model that describes 
the charge ordering transition and the coupling of intra-rung 
charge fluctuations to the spins. Sec. III is devoted to the 
description of the numerical methods employed. Sec. IV gives 
the computational results for the correlation functions and 
the conjectured phase diagram of CO, furthermore the influence 
of the spin exchange on the critical CO parameter is studied. 
In addition we investigate the 
finite size scaling of the spin excitation gap and its dependence 
on the degree of charge order. 
Moreover we have studied the effect of spin exchange bond-alternation 
on the energy gap and the scaling of spin gap for small value of
dimerization $\gamma$.  
A summary of our results is given in Sec. V.


\section{Charge ordering and Pseudo-spin models} 

An attractive feature of the \Va compound that simplifies its theoretical 
study, is the fact that the charge ordering happens to be in an already
insulating mixed valent (MV) state. It is characterized by a noninteger
valence 4.5+ of each vanadium atom of the rung which therefore is occupied
by only one d- electron on the average. The ordering is due to 
the fact that the large intra-rung hopping ${\rm t}_a$ 
leads to a half-filled (rung-) {\em bonding} band that is separated 
into an (empty) upper and (filled) lower Hubbard band. The 
effective insulating gap is then a charge transfer gap approximately 
given by ${\rm t}_a$.
Although this scenario is suggestive, a detailed theory for the 
insulating state of \Va is still lacking. Due to the charge gap,
each rung accommodates only one d-electron which can occupy the right (R) 
or left (L) position of the rung. 
As first proposed in Ref.({\onlinecite{Thalmeier1}), this charge 
degree of freedom can be described by an Ising pseudospin variable
$\tau_z$ and its resonating behaviour
by the transverse isospin variable $\tau_x$. The latter is responsible
for the tendency towards  the formation of a bonding state.
The Coulomb interactions between d-electrons on different rungs 
may then be mapped onto Ising-like  interactions on the one-particle 
subspace of the rungs. Altogether the charge degrees of freedom are 
then described by

\begin{eqnarray}
H_c= -{\rm t}_a\sum_{i}\tau_x^i
+ {\rm K}\sum_{\ll i,j\gg}\tau_z^i\tau_z^j 
-{\rm K}'\sum_{\langle i,j\rangle}\tau_z^i\tau_z^j 
\label{Hc}
\end{eqnarray}

\begin{figure}
\epsfxsize=8cm \epsfysize=7cm  \epsffile{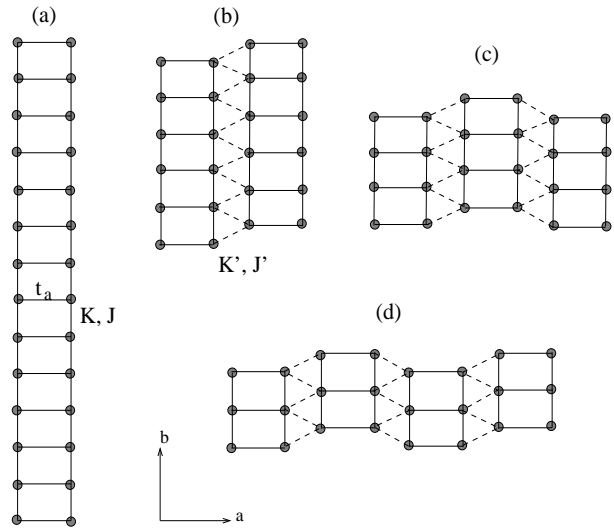}
\vspace{0.5cm}
\caption[]{ 
Schematic representation of the crystal structure of a vanadium layer
of \Va. The vanadium ions are denoted by filled circles.
The couplings in  Eqs.(\ref{Hc} \& \ref{hcs}) are also shown:
the hopping amplitude (${\rm t}_a$) on each rung, effective pesudo-spin $K$
and spin $J$ couplings on each ladder and between ladders $K', J'$. 
Different configurations that we have considered in exact
diagonalization are denoted by: (a) A single ladder is
considered with $N=28$. (b) Two interacting ladders via $K', J'$ compose a
lattice of $N=24$. (c) Three interacting ladders and in (d) four interacting
ladders.
}
\label{fig1} 
\end{figure}

The first term describes the intra-rung hopping, which is the largest 
hopping element ${\rm t}_a$, see Ref.(\onlinecite{Yaresko}). 
The terms $\sim {\rm K}, {\rm K}'$ are effective 
intra- and inter-ladder Coulomb interactions, 
respectively\cite{Thalmeier1} (Fig.\ref{fig1}),
which contain also the effect of virtual inter-rung hopping.
An interesting particular case of this Hamiltonian occurs 
if we consider only a single ladder 
(i.e. set ${\rm K}'=0$). Then this Hamiltonian describes the 1D Ising model 
in a transverse field of strength ${\rm t}_a$ whose exact solution, 
including the correlation functions $\langle \tau_z^i\tau_z^j \rangle$ and 
$\langle \tau_x^i\tau_x^j \rangle$ (j=i$\pm$1), is known \cite{Pfeuty}.
It has been discussed in the present context in Ref.(\onlinecite{Yushankhai}). 
The model is able to describe the appearance of the zig-zag charge ordering 
along the ladder as a function of the control parameter K/t$_a$;
strictly speaking an infinitesimal staggered field due to inter-ladder 
coupling is necessary to stabilize the order parameter\cite{Yushankhai}. 
Due to the peculiar structure of the Trellis lattice (rungs of 
neighbouring ladders are shifted by half a lattice constant b) 
the inter-ladder effective Coulomb interaction ${\rm K'}$ leads to a 
frustration in the zig-zag CO state. If it exceeds a critical 
value ${\rm K}_c'$, then  the zig-zag CO is destroyed 
and in-line CO would be prefered. 
This transition corresponds to a change in the modulation vector of the 
pseudo-spin density $\tau_z(\vec{Q}$) from 
$\vec{Q}=(\frac{\pi}{a},\frac{\pi}{b})$ 
to $\vec{Q}=0$. The Hamiltonian of Eq.(\ref{Hc}) with K$>$0 describes a 
2D frustrated Ising model in a transverse field (ITF) on a trigonal lattice.

In  this work our aim is to present a more detailed  study 
of the correlation functions and phase 
diagram of this model.\\

It was shown in Ref.(\onlinecite{Yushankhai}) that the spin exchange 
between the ladders is strongly influenced by the charge fluctuations 
and the charge order parameter. In this investigation the intra-rung 
charge fluctuations have been 
eliminated by using the 1D- Ising model exact solution.
This is only possible for one ladder and not in the 2D frustrated 
Ising model appropriate for the Trellis lattice. Therefore we will 
keep the general interaction term of charge and spin degrees of 
freedom as proposed in Ref.(\onlinecite{Thalmeier1})

\begin{eqnarray}
H_{CS}
&=&J\sum_{\ll i,j\gg}\frac{1}{16}[(1+\tau_z^i)(1+\tau_z^j)
   \vec{\sigma}^i\vec{\sigma}^j \nonumber\\
   &+&(1-\tau_z^i)(1-\tau_z^j)
   \vec{\sigma}^i\vec{\sigma}^j] \nonumber\\
   &+&J'\sum_{\langle i,j\rangle}\frac{1}{32}[(1+\tau_z^i)(1-\tau_z^j)
   \vec{\sigma}^i\vec{\sigma}^j \nonumber\\
   &+&(1-\tau_z^i)(1+\tau_z^j)
   \vec{\sigma}^i\vec{\sigma}^j]
\label{hcs}
\end{eqnarray}

The first and second term describe the coupling of charge fluctuations 
to the intra- and inter-ladder exchange, respectively. For clarity we use
explicitely the R,L pseudospin projectors $\frac{1}{2}(1\pm\tau_z)$ in the 
Hamiltonian. Approximate 
expressions for the superexchange constants $J,J'$ have been given 
in Ref.(\onlinecite{Thalmeier1}). In general we ignore superexchange 
paths along the ladder diagonals, although this is not completely justified 
\cite{Yaresko}; we will consider its effects in a few cases. 
So far we have introduced a large number of coupling constants as free 
parameters in these models. It is not our purpose to compute the properties
of them for any range of the parameters involved.
Instead, we shall be interested in those values that are physically relevant
for the underlying \Va compound. They were obtained in LDA+U 
calculations\cite{Yaresko} assuming full charge order. 
In this method, the 3d-occupation of each vanadium site 
is fixed and an empirical on- site Coulomb interaction U is added
to the LDA potential which shifts occupied 3d-states downward 
and unoccupied 3d-states upward with respect to the Fermi level 
and in this way leads to a
proper insulating state for \Va. From these LDA+U calculations\cite{Yaresko}
the following hopping and exchange constants were obtained: 
${\rm t}_a$= -340 meV, $J=$ 30 meV and $J'=$ -5 meV, i.e. there is 
antiferromagnetic (AF) intra-ladder and ferromagnetic (FM) 
inter-ladder exchange. As the effective inter-site Coulomb constants are not 
known, they are left as variable parameters in a range of up to 1 eV. 
The on-site Coulomb interaction, which does no longer appear in our 
effective low energy Hamiltonian, has the magnitude $U= 4$ eV.

\section{Ladder Hilbert space and computational technique}

In our numerical computation,
each rung {\it i} is associated with the four dimensional 
Hilbert space \{$|\sigma_i\rangle\otimes|\tau_i\rangle$\} where $\sigma_i, 
\tau_i =\uparrow, \downarrow$. We classify the states used according to 
their total spin quantum number $\sigma_z$ and restrict to the space with  
$\sigma_z$=0 for the ground state calculations. Note that the total 
pseudo spin z-component $\tau_z$= 0 is not conserved due to the first 
term in Eq.(\ref{Hc}). As a consequence of this, 
all states in the pseudo-spin sector have to be 
included in the calculation. Therefore the maximum number of sites which 
we treat is N=28. They may be arranged in 
different structures, some 
of them are shown in Fig.\ref{fig1}. 
In Fig.(\ref{fig1}-a) we have plotted an isolated single ladder where N=28.
The intra-ladder coupling constant K (which is the result of Coulomb 
interaction on neighbouring rungs) exists on each leg of the ladder. 
Similarly the 
exchange constant $J$ on each leg represents the coupling of the
spin degrees of freedom of each electron. The hopping parameter t$_a$
is defined on each rung. Most of our calculation on the single ladder
is considered with periodic boundary condition (BC) in the legs direction.
We have only considered the open BC when we compare the correlation functions
of a single ladder and two coupled ladders.
The interactions of two ladders is assigned in Fig.(\ref{fig1}-b) with
inter-ladder couplings K$'$  and $J'$ which is denoted by dashed lines.
All of results presented here on two coupled ladders are in open BC. The 
interaction between three and four ladders is depicted in 
Fig.(\ref{fig1}-c and d). The results of computations on this latter
configurations are not presented here because the lattice dimension on 
each direction is very small and we cannot avoid boundary effects.
However this results help us to see what will be the effect of more
ladder interation and  find a qualtitative phase diagram for 
our model.

\noindent We use a single ladder to study the finite 
size scaling of various quantities as function of the number of sites N. 
We also use some more two dimensional structures as shown in Fig.\ref{fig1} 
corresponding to various N to investigate the effects of inter-rung 
interactions, especially concerning the phase diagram as a function of 
${\rm K}, {\rm K}'$. 
We represent  correlations for open boundary 
conditions on  the ladder since for the former the 
correlation functions are symmetric around the midpoint of the ladder 
and therefore presumably contain less information. 
The periodic boundary condition is applied in the b-direction and
used to compute the energy levels.   


\section{Results of exact diagonalization}

In this section we present the results of our calculations for 
correlation functions of the Hamiltonian ($H$) with both spin and pseudo-spin
degrees of freedom ($H=H_c+H_{CS}$).
From this information, we are able to  conjecture a  phase 
diagram in the $(K,K')$- plane. Furthermore, we also study 
the influence of the exchange 
on the critical value $K_c$ for charge ordering in the single ladder. 
Subsequently, we come to the important question about the nature of excited 
states (charge type vs. spin type) and discuss the finite size scaling of the
energy gap in a single ladder. 
We have also considered the effect of bond-alternation (dimerization, $\delta$)
of the spin exchange on the single-ladder system. This shows the scaling 
behaviour of spin gap due to $\gamma$. 

\begin{figure}
\epsfxsize=8cm \epsfysize=7cm  \epsffile{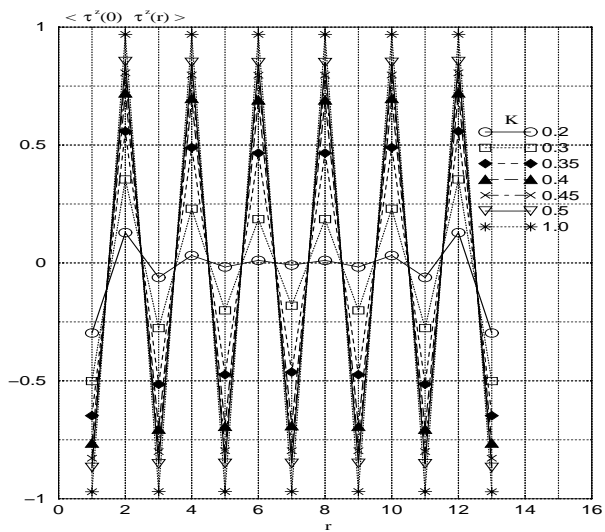}
\vspace{0.25cm}
\caption{Pseudospin correlation function 
($\langle \tau^z(0) \tau^z(r) \rangle$)
for a single ladder (Fig.(\ref{fig1}-a)) versus $r$, the distance between
rungs. Periodic boundary condition are considered along legs of ladder so 
that the correlations are symmetric around the mid-point. 
By increasing $K$,  an evolution
from a disordered phase (no-long-range order) at small value 
of K ($\le 0.35$) to long-range order happens 
for large value of K (J=30 $meV$).}
\label{fig2} 
\end{figure}

\subsection{Correlation functions and CO transition}

The pseudo-spin interaction ${\rm K}>0$ along the ladder legs is of 
``antiferro" type because physically the d-electrons sitting on the 
same side of adjacent rungs have a larger repulsion than if they 
were located across the ladder diagonal. Thus the second term in 
Eq.(\ref{Hc}) tends to localize the electrons on opposite sites of the 
ladder in zig-zag fashion (Fig.\ref{fig1}). On the other hand the first 
term in Eq.(\ref{Hc}) which describes the hopping energy within the 
rung prefers the disordered mixed valent state where the d-electron 
occupies the symmetric rung-bonding state with equal probability of 
sitting on the L and R sites. This state is realized for ${\rm K}=0$ 
(and ${\rm t}_a$ as given above from LDA+U calculations). 
If K increases above a  critical value ${\rm K}_c={\rm t}_a$,  
eventually the ordered state with a zig-zag correlation is preferred. This 
behaviour is clearly exhibited in Fig.\ref{fig2} for the single ladder (K$'$=0).

The correlation function for a single ladder with periodic BC is plotted
for N=28 in Fig.(\ref{fig2}). For small values of K (K$<0.4$) there is
no LRO and correlations decay very fast. By increasing K (K$>0.4$), 
the correlation function shows zig-zag (alternating) LRO where the 
amplitude of the correlations is approximately constant for all distances (r).
The critical interaction can be seen to be ${\rm K}_c$= 0.35-0.4 eV 
$\simeq$ ${\rm t}_a$. 
The coupling to adjacent ladders (third term in Eq.(\ref{Hc})) would however 
prefer the ``ferro'' type arrangement, which leads to a frustration of 
charge (pseudo-spin) order. The effect of this frustration is clearly 
visible in Fig.\ref{fig3} and Fig.\ref{fig4} 
for the correlation functions with K$'>$0, most 
directly for the 2$^{nd}$ neighbour rung (r=2). 
Its correlation for K$'=0$ is 
of ferro type, then increasing K$'$ becomes first antiferro-type and 
finally, when K$'$ is large enough so that the in-line ordering is prefered, 
it becomes of ferro type again. This non-monotonic behaviour of 
$\langle\tau^z(i)\tau^z(i+2)\rangle$ is due to the competition 
of different interaction paths, either directly along the legs 
of the given ladder, involving K, or indirectly via adjacent ladders 
which involves K$'$.

A similar behaviour is seen in the two-dimensional lattice by considering 
three or four ladders interacting via K$'$. In this case the number of rungs
is reduced to four and three respectively in each ladder 
(Fig.\ref{fig1}-(c) and (d)). 
We shall  explain the case
of the two-ladder system in more detail. In this case the system is composed of
two 6-rung ladders with open boundary conditions imposed. The pseudo-spin
correlation functions are plotted in Fig.\ref{fig3} and Fig.\ref{fig4}. 
We have plotted
$\langle \tau^z(0) \tau^z(r) \rangle$ along the b-direction for both
single and double ladder systems. The case of a single-ladder is added for
comparison. In Fig.\ref{fig3}, the value of K is fixed to K=0.35 $<$ K$_c$. 
The case of single-ladder (K$'$=0) shows a rapid decay with distance which 
verifies 
that no LRO exists in the ladder. By turning on 
the K$'$ coupling, the 
second ladder interacts with the first one. We have plotted the correlation 
functions at the same points of the  single-ladder for different value
of K$'$. When K$'$=0.175 or 0.35 eV, the qualitative behaviour does not change 
which verifies the absence of LRO. By increasing K' the ordering of 
the first neighbour
(r=1) goes from anti-ferro to ferro type. For the 2$^{nd}$ neighbour (r=2)
this ordering goes from ferro- to anti-ferro and finally again to ferro- type
for enough high value of K$'$ (in this case K$'>1.25$).  
As mentioned above, the competition between different paths of interactions
via K and K$'$ is responsible for this behaviour. It happens for 
0.35 $<$ K$'<$ 0.7 eV 
which is a phase with no LRO and 
will appear in the 
phase diagram (Fig.\ref{fig5}). 
For high value of K$'$ an in-line LRO appears on each ladder of the model.

\begin{figure}
\epsfxsize=8cm \epsfysize=7cm  \epsffile{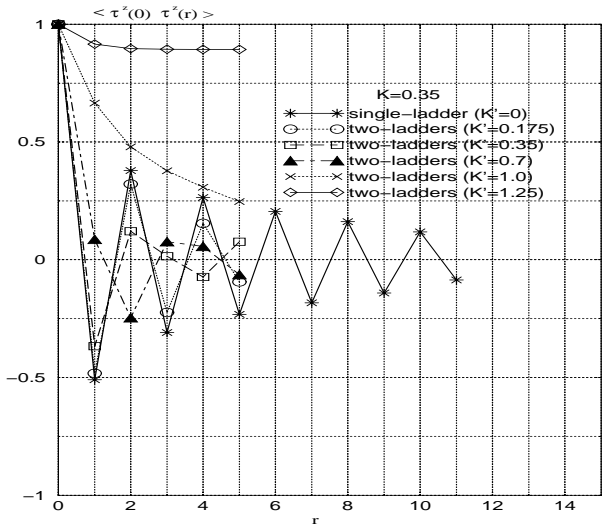}
\vspace{0.2cm}
\caption{Pseudospin correlation function 
($\langle \tau^z(0) \tau^z(r) \rangle$)
for both single (Fig.(\ref{fig1}-a)) and double ladder (Fig.(\ref{fig1}-b))
versus distance ($r$). In both cases the intra-ladder coupling (K) is fixed
to K=$0.35 < K_c$. Then the correlations decay very fast and do not show 
long-range zig-zag ordering. By increasing K$'$ (inter-ladder coupling) 
deviation from single ladder results start and leads to a disorderd regime for
an intermediate value of K$'$=0.35. For sufficient size of K$'$ an in-line
ordering will appear. (Correlations computed with open 
boundary condition, J=30 $meV$ and J$'$=-5 $meV$).}
\label{fig3} 
\end{figure}

In addition, Fig.\ref{fig4} 
shows a similar behaviour 
like Fig.\ref{fig3}, while K is fixed to 
K=0.5 eV $>$K$_c$ in the zig-zag ordered phase of a single-ladder system.
The correlation function
shows that for the single-ladder (K$'$=0), a true zig-zag LRO 
exists. As before, a small value of K$'$ does not change the ordering  of 
the single-ladder (K$'$=0.25 and 0.5 eV). For an intermediate value of K$'$,
the disordered (no LRO) phase will appear (0.5 $<$ K$'<$ 1.5 eV).
Then by increasing K$'$ a complete in-line ordering starts to stabilise.

\subsection{Phase Diagram for CO}

As we have seen in the previous sections,
the transition from antiferromagnetic (zig-zag) to 
ferromagnetic  (in-line) type CO depends on the ratio of K to ${\rm t}_a$ 
and on the ``frustration'' ratio K$'$/K. As the 2$^{nd}$ neighbour (r=2) 
correlation along the ladder shows the transition is not abrupt 
but gradual. 
This signals an intermediate region in the K, K$'$ plane which has 
to be considered as a phase of no LRO. We believe that this phase 
is very similar to the ``floating phase'' discovered in the 2D-ANNNI 
(anisotropic next nearest-neighbour Ising) model on a square lattice even 
though in this case it exists only at 
finite temperature (e.g. Ref.(\onlinecite{Bak})). 
The inherent geometric frustration of the Trellis 
lattice might stabilize it even for zero temperature. The phase diagram 
obtained from systematic study of the correlation 
function is shown in Fig.\ref{fig5}. 
For simplicity, let us neglect the spin coupling and consider Eq.(\ref{Hc})
in two marginal cases. First, if K$'=0$, Eq.(\ref{Hc}) reduces to a
one-dimensional ITF model. Then for K$<$K$_c$ there is 
no LRO for charge degrees of freedom while a 2$^{nd}$-order quantum 
phase transition
will occur at K$_c$ to an anti-ferroelectric (zig-zag) LRO. 
In the other extreme case where K=0, Eq.(\ref{Hc}) is reduced to 
a two-dimensional
ITF model on a square lattice. Obviously for small value of K$'$, the first 
term in Eq.(\ref{Hc}) prevents the existence of LRO for 
charge degrees of freedom. If K$'$ goes to large values respect to t$_a$,
the two-dimensional square lattice shows a ferro-electric (in-line) ordering.
Therefore the phase diagram of Eq.(\ref{Hc}) in the K-K$'$ plane consists
of two boundary lines which divide a phase with no LRO in the intermediate 
regime from the other in-line and zig-zag LRO. This phase boundary is an
extension of two critical points K$_c$ and K$'_c$  in the K-K$'$ plane which 
is supported by numerical data on two-dimensional clusters 
(Figs.\ref{fig3} and \ref{fig4}).

\begin{figure}
\epsfxsize=8cm \epsfysize=7cm  \epsffile{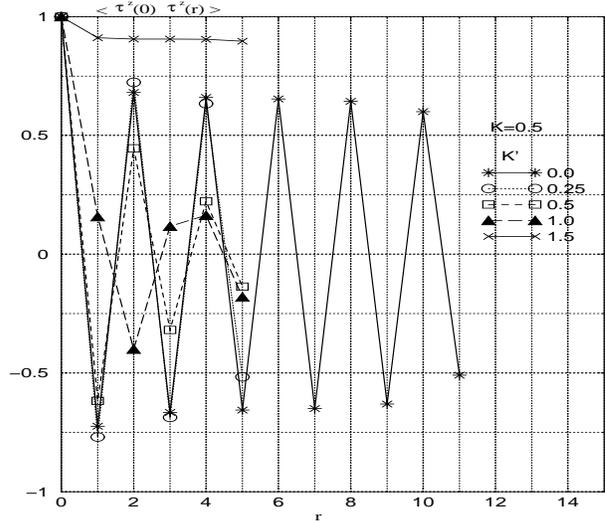}
\vspace{0.2cm}
\caption{This plot is similar to Fig.(\ref{fig3}) but for different values 
of K and K$'$. Here K is fixed to K=$0.5 > K_c$ which shows long-range zig-zag 
order on the single ladder (K$'$=0). By turning on the inter-ladder coupling
(K$'$), no qualitative changes 
are observed for small value of K$'$=0.25.
By increasing K$'$, the system goes through a disordered phase for intermediate
values of K$'$=1.0 and then enters to the in-line ordered phase for large 
value of K$'$=1.5. (Correlations are computed with open boundary 
condition J=30 $meV$ and J$'$=-5 $meV$).}
\label{fig4} 
\end{figure}

The phase boundary for the upper in-line phase and the lower zig-zag 
phase are obtained by requiring a threshold value of 90\% for the 
correlation of each type. As critical values we obtain K$_c\simeq$ 0.39 eV
(the effect of spin coupling has been considered) 
and K$'_c$ is the critical value of a two-dimensional ITF model
on a square lattice including the spin coupling by Eq.(\ref{hcs}).
If we neglect the effect of  Eq.(\ref{hcs}) 
then K$'_c\simeq$0.054 eV \cite{Pfeuty2}.
These phase boundaries enclose the disordered 
phase with no clearly discernible long range correlation.
Although specifying the properties of the intermediate no LRO phase is
beyond our numerical computations our data show that the period of 
oscillation of correlation functions has doubled in comparison to the zig-zag
ordered phase. This can be seen in Fig.(\ref{fig3}) at K$'$=0.7
and in Fig.(\ref{fig4}) at K$'$=1.0.

\begin{figure}
\epsfxsize=8cm \epsfysize=7cm  \epsffile{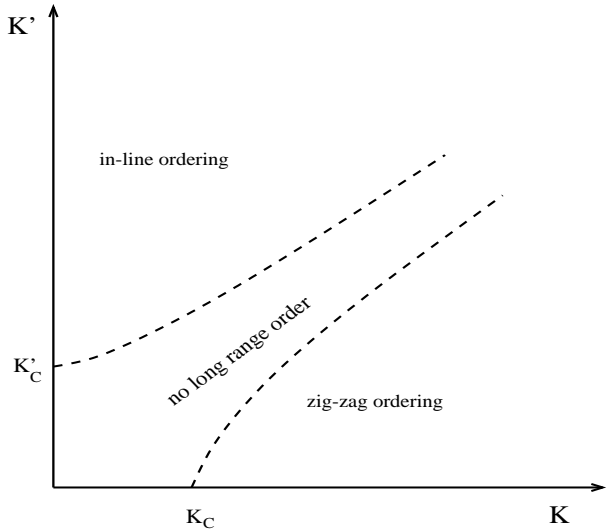}
\vspace{0.2cm}
\caption{Charge-order (CO) phase diagram in the space 
of intra-(K) and inter-ladder (K$'$)
couplings. K$_c$ represent the critical value of K in a single ladder
configuration (Fig.(\ref{fig1}-a)). On the vertical axis K$'_c$ is the 
critical value of a two-dimensional (ITF + spin exchange effect) on a square
lattice. The dashsed lines show the qualitative phase boundary between three
different phases where zig-zag ordering happens in the lower-right part and
in-line ordering in the upper-left. A phase with no LRO separates the two 
ordered phases.}
\label{fig5} 
\end{figure}

\noindent The phase boundaries for CO also depend on the size of 
the exchange constants.
This has been studied for the point (K,K$'$)=(K$_c$,0), corresponding to 
the single ladder case in  more detail as shown in Fig.\ref{fig6}. The 
critical value K$_c$ for development of long range order is defined 
by the simultaneous crossing point where curves for different ladder 
lengths (N=20-28) meet. One can clearly see that K$_c$ shifts as a 
function of the exchange $J$ along the ladder leg. This can be explained 
as follows. In the disordered phase there is a certain amount of
stabilisation energy of the spin singlet. The incipient zig-zag CO 
which localizes the d-electrons on opposite sites of the ladder 
diagonal tends to reduce this stabilisation energy and hence, if J 
increases the critical K$_c$ for CO to set in also has to increase. 
It is shown for two different values of $J$ (30, 100 meV) in Fig.\ref{fig6}.
In the case of single-ladder  by setting $J=0$, we have a one-dimensional
ITF model where K$_c$=t$_a$. 
It is shown in Fig.\ref{fig6}-a that the intersection
of the pseudo-spin correlation function for different values
of N (20, 24, 28) is at K$_c$=0.39 eV where $J$=0.03 eV (t$_a$=0.34 eV).
A similar curve in Fig.\ref{fig6}-b shows that the intersection is at K$_c$=0.40 eV
where J is increased to $J$=0.1 eV at fixed t$_a$. This verifies the monotonic
dependence of K$_c$ on the exchange constant $J$.

\begin{figure}
\epsfxsize=8cm \epsfysize=8cm  \epsffile{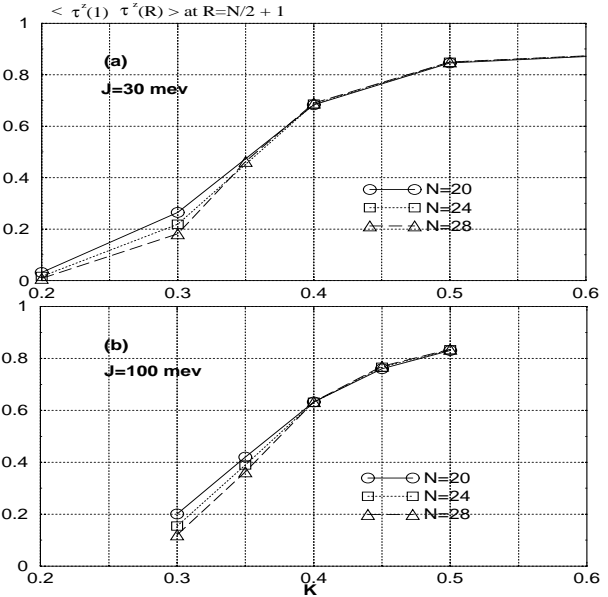}
\vspace{0.2cm}
\caption{The pseudo-spin correlation amplitude 
at fixed distance R=$\frac{N}{2}$ versus
the intra-ladder coupling (K) for single ladder. Different sizes of
ladder (N=20, 24, 28) are implemented to 
track the crossing point of curves,
which is the transition point. (a) The effect of spin exchange (J) is 
considered with the value of J=30 $meV$ then the crossing point happens
at K$_c\simeq$0.39. (b) In this case J=100 $meV$ which causes an increase
to the value of K$_c\simeq$=0.4. (Periodic boundary conditions are considered
along the legs of ladder).}
\label{fig6} 
\end{figure}

\subsection{Nature of excited states, 
finite size scaling and effective exchange}

We have implemented the Lanczos method in different $S^z_{tot}$ subspaces
to determine the structure of excited states.
For the single-ladder system, the ground state and first excited state are 
always spin singlet and triplet state respectively, as long as we do not 
include the exchange path along the ladder diagonals. This is 
true even  
in the coupled ladder system (2D Trellis lattice). 
As  we have a Heisenberg exchange term, the 
singlet- triplet gap has to vanish in the thermodynamic limit 
N$\rightarrow\infty$.

However, from the finite size scaling behaviour 
we may extract the ``effective exchange" along the ladder for a pure spin
model, 
which is modified from its bare value J due to the effect of 
intra-rung charge fluctuations. In Fig.\ref{fig7}-a the scaling behaviour 
of the singlet- triplet gap $\Delta$ for a single ladder is shown 
as a function of 1/N (N=16, 20, 24, 28) for various interaction paramters K. 
All curves show a decrease in the energy gap with increasing N, which
leads to gapless behaviour by extrapolation to $N \rightarrow \infty$.
We use a scaling form for the energy gap given by

\begin{equation}
\Delta(N)\simeq\Delta_\infty +\frac{\alpha}{N^\nu}
\end{equation}

where $\Delta_\infty\rightarrow$0 for N$\rightarrow \infty$, $\alpha$ is 
constant and $\nu$ a scaling exponent. These values are given 
in Table 1 for various K and we see that $\nu$ $<$ 1 for K $<$ K$_c$ 
and  $\nu$ $>$ 1 for K $>$ K$_c$. 

\begin{center}
\begin{tabular}{|c|c|c|} 
\hline
$K$   &  $\alpha$   & $\nu$   \\ 
\hline
$0.3$  & $ 0.03379$ & $ 0.75224$  \\ 
\hline
$0.35$   & 0.03332   & 0.88581   \\ 
\hline
$0.392$  & $0.03633$  & $1.04005$   \\
\hline
$0.4$   & 0.03683   & 1.06949   \\ 
\hline
$0.45$  & $ 0.03699$ & $ 1.21717$  \\ 
\hline
$0.5$  & $ 0.02862$ & $ 1.23286$  \\ 
\hline
\end{tabular}
\end{center}
Table 1. 
Numerical values of the coefficients in the scaling function 
of the energy gap in Fig.\ref{fig7}-a. The scaling exponent is $\nu < 1$
for $K<K_c \simeq0.39$ and $\nu > 1$ for $K>K_c$. The scaling form is 
$\Delta= \Delta_{\infty} + \frac{\alpha}{N^{\nu}}$. (Data has been
obtained by imposing a least square root error of $10^{-8}$ 
for the scaling function ).

\begin{figure}
\epsfxsize=8cm \epsfysize=8cm  \epsffile{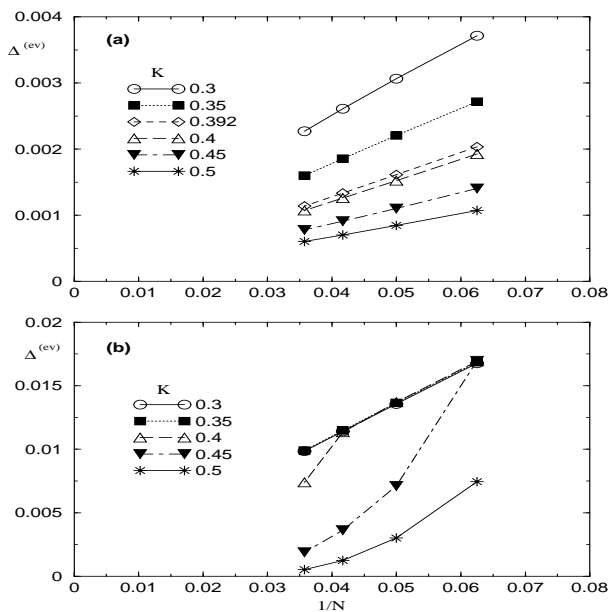}
\vspace{0.2cm}
\caption{
The energy gap to the 1$^{\rm st}$
excited state versus 1/N of the
single ladder (Fig.(\ref{fig1}-a)) for different values of intra-ladder
coupling (K). (a) All data show gapless behaviour in the
$N\rightarrow \infty$. The spin exchange is J=30 $meV$.
(b) In addition to J  we have added  diagonal exchange terms 
(J$_d$= 32.7 $meV$) on each plaquette of ladder. 
For K$<$K$_c$=0.39, we observe qualitative behaviour similar to part (a) 
where the first excitation
is a spin triplet. But for K$>$K$_c$ the behaviour is completely different
because the excitations are charge type. The excited state has the
same total spin $S_{tot}$=0.
}
\label{fig7} 
\end{figure}

This shows the different curvature of $\Delta$ versus $1/N$.
(To have the best fit of the data in Fig.\ref{fig7}-a 
at fixed $\nu=1$ we have
considered more terms in the curve fitting form;
$\Delta(N)=\Delta_{\infty}+\frac{C_1}{N}+\frac{C_3}{N^3}+\frac{C_5}{N^5}$ which 
is justified for the case of pure spin- $\frac{1}{2}$ 
chain \cite{Auerbach}).
For a pure spin-$\frac{1}{2}$ system 
without coupling to charge degrees of freedom, one has $\nu$=1 and 
the coefficient $C_1$ is related to

\begin{equation}
C_1=2\pi^2S^2J_{{\rm eff}}
\end{equation}

\noindent Comparison with our $C_1$ coefficient from 
the scaling plots in Fig.\ref{fig7}-a 
gives us the variation of the effective exchange constant J$_{{\rm eff}}$(K) 
with the interaction K that controls the charge order (for a given t$_a$). 
This variation is shown in Fig.\ref{fig8}. 
A steep drop in J$_{eff}$(K) is seen 
in the region where the charge order sets in (K$\simeq$0.39). Eventually 
for K$\gg$t$_a$ when zig-zag CO is almost complete, J$_{eff}$ has to 
approach zero because in the zig-zag structure the exchange J along 
the legs is ineffective.\\

The situation changes when we also include exchange along the ladder 
diagonals. 
In this case we will add an exchange term similar to the first
term of Eq.(\ref{hcs}) along the ladder diagonal in each plaquette (not 
shown in Fig.1).
The value of diagonal exchange is considered to be $J_d= 32.7meV $ 
which has been obtained by LDA+U calculations in Ref.(\onlinecite{Yaresko}).
For K $<$ K$_c$ the first excited state is a spin triplet, 
for K $>$ K$_c$ however 
the first excited state is not a spin triplet, rather it has spin zero like
the ground state and therefore it should be interpreted as a charge 
excitation. This leads 
to a very different scaling of the energy gap for K $>$ K$_c$ 
(Fig.\ref{fig7}-b) while 
the case K $<$ K$_c$ is very similar to the model without diagonal 
exchange (Fig.\ref{fig7}-a). 
The scaling behaviour in Fig.\ref{fig7}-b for K$<$0.35 is roughly linear which 
behaves like the former spin excitations while for K$>$0.4 the scaling
behaviour is completely different.

\begin{figure}
\epsfxsize=8cm \epsfysize=7cm  \epsffile{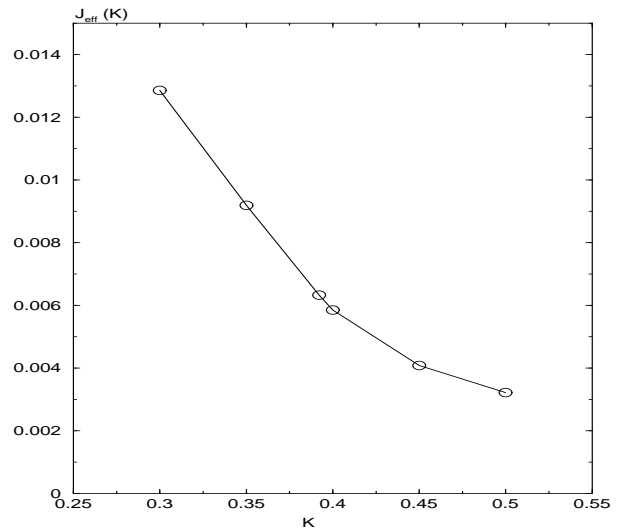}
\vspace{0.2cm}
\caption{Effective superexchange coupling versus K. It shows a steep decrease 
close to the transition point K$_c$=0.39. Moreover the effective exchange 
goes to zero in the completly zig-zag orderd phase (K$\rightarrow \infty$).}
\label{fig8} 
\end{figure}

We have also considered the effect of dimerization ($\delta$) of the
exchange coupling on the energy gap of a single-ladder. This is of interest
because it seems now clear that the charge disordered ladders in the low
temperature structure of \Va discussed in the introduction have a large
superexchange dimerisation due to the shift of oxygen 
positions\cite{Yaresko}. Therefore we consider the influence of the 
dimerisation for K$<$K$_c$ in the disordered regime. This means that 
the exchange constant (J) of the first term in Eq.(\ref{hcs}) is replaced
by $J(1 \pm \gamma)$ in an alternating way. The dimerization parameter varies as
$\gamma$=0.01, 0.05, 0.1, 0.2, $\dots$, 0.6. Obviously an energy gap opens by
turning on $\gamma$ which does not scale to zero 
with increasing N ($N \rightarrow \infty$) as it 
did for the undimerized case $\gamma$=0. 
In Fig.\ref{fig9}-b we have plotted the log-log plot of energy gap versus 
$\gamma$.
For small value of $\gamma$ the energy gap scales as 
$\Delta \sim \delta^{\eta}$ where $\eta=2.25 \pm 0.25$.

\begin{figure}
\epsfxsize=8cm \epsfysize=8cm  \epsffile{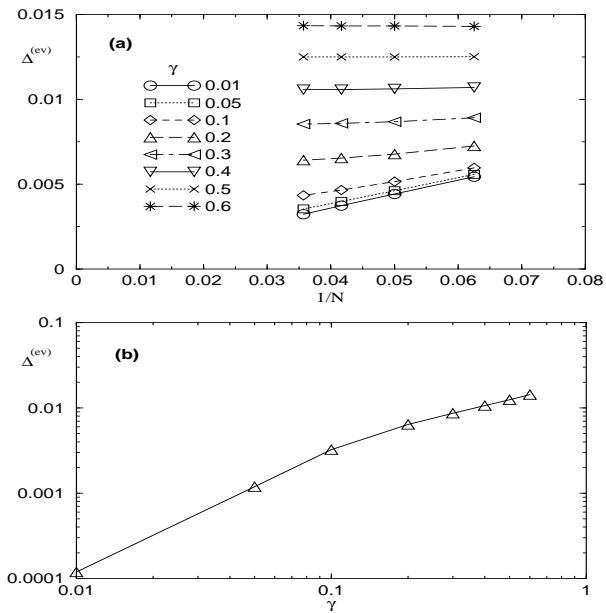}
\vspace{0.2cm}
\caption{Spin gap energy versus 1/N in the dimerized 
single-ladder for K=0.2$<$K$_c$ and J=30 $meV$.
(a) Data are plotted for different value of dimerization parameter
($\gamma=0.01, \dots, 0.6$). All  curves show up a finite gap in 
the $N\rightarrow \infty$
(for $\gamma > 0.05$ it is more apparent).
(b) The extrapolated energy gap is plotted versus $r$ in a log-log scale.
For small value of $\gamma$ ($\gamma <0.1$) we observe the scaling behaviour,
$\Delta=\gamma^{\eta}$.}
\label{fig9} 
\end{figure}


\section{Summary}
The main results of our exact diagonalisation investigation for 
the present \Va model are the following: For the single ladder the 
ratio of K/t$_a$ controls the onset of 
charge order which is observed in the evolution of the pseudospin 
correlation function (Fig.\ref{fig2}). 
This is similar as in the exact 1D Ising 
model in transverse field results \cite{Pfeuty}.

\noindent For the 2D Trellis lattice, it depends on the relative size of 
intra- and inter-ladder effective Coulomb interactions K,K$'$ and 
the size of K with respect to the intra-rung hopping t$_a$ whether 
in-line or zig zag structures are realized. This agrees with LDA+U 
total energy calculations which show that these two structures are 
very close in energy with the zig-zag structure slightly lower \cite{Yaresko}.
In addition however we have shown that in a wide region in the K, K$'$ 
plane a disordered phase should exist which does not show long range 
correlations of charge due to the inherent frustration of the Trellis 
lattice structure. According to L\"udecke et al \cite{Luedecke} in 
the real \Va structure only every second ladder is charge ordered and 
the intervening ladders stay in the MV state. This might be an 
indication that \Va is not far from the regime with a disordered state,
i.e. close to a quantum critical point.

\noindent We have also studied the effect of spin coupling on the charge 
order transition. We have shown that K$_c$ has a monotonic dependence on the
exchange constant($J$), i.e. for J=30 meV it leads to 15 percent increase
of the 1D ITF result.

\noindent We have addressed the issue of excitation type in our charge-spin
model. As long as the spin exchange along the ladder diagonal is absent the 
excitations are singlet to triplet. By adding a diagonal spin exchange on 
ladders we will observe charge excitations in the ordered phase while 
it is still spin type in the disordered phase.

From the finite size scaling analysis of the singlet triplet excitation 
gap we obtain the dependence of the effective inter-rung exchange on the 
ladder as function of the degree of charge order, i.e. as function of K. 
A steep drop of J$_{eff}$ is observed around the critical value K$_c$ for 
charge order. Eventually it will be reduced to zero in the fully 
CO regime in qualitative agreement with perturbation calculations
 \cite{Yushankhai} 

\section{Acknowledgement}
A.L. would like to thank A. Bernert, V. Yushankhai and I. Peschel for
fruitful discussions and acknowledge 
Max-Planck-Institute for the Physics of Complex Systems for computer 
facilities and M.A.M.D. would like to thank
the Centro de Supercomputaci\'on Complutense for the allocation
of CPU time in the SG-Origin 2000 Parallel Computer.
M.A.M.-D. was supported by the DGES spanish grant
PB97-1190.

\end{document}